\documentclass[conference,a4paper]{APSIPA2025}
\usepackage{amsmath}
\usepackage{bm}
\usepackage{graphicx}
\usepackage{multirow}
\usepackage{threeparttable}
\usepackage[backend=biber,style=ieee,]{biblatex}
\usepackage{geometry}
\usepackage{fancyhdr}

\fancypagestyle{firststyle}{
  \fancyhf{}
  \fancyhead[C]{2025 Asia Pacific Signal and Information Processing Association Annual Summit and Conference (APSIPA ASC)}
}

\begin{document}

\title{Voice Privacy Preservation with Multiple Random Orthogonal Secret Keys: Attack Resistance Analysis}

\author{
\authorblockN{
Kohei Tanaka\authorrefmark{1}, Hitoshi Kiya\authorrefmark{1} and Sayaka Shiota\authorrefmark{1}
}

\authorblockA{
\authorrefmark{1}
Tokyo Metropolitan University, Japan
}



}

\maketitle
\thispagestyle{firststyle}
\pagestyle{fancy}
\pagestyle{empty}

\begin{abstract}
Recently, opportunities to transmit speech data to deep learning models executed in the cloud have increased. This has led to growing concerns about speech privacy, including both speaker-specific information and the linguistic content of utterances. As an approach to preserving speech privacy, a speech privacy-preserving method based on encryption using a secret key with a random orthogonal matrix has been proposed. This method enables cloud-based model inference while concealing both the speech content and the speaker identity. However, the method has limited attack resistance and is constrained in terms of the deep learning models to which the encryption can be applied. In this work, we propose a method that enhances the attack resistance of the conventional speech privacy-preserving technique by employing multiple random orthogonal matrices as secret keys. We also introduce approaches to relax the model constraints, enabling the application of our method to a broader range of deep learning models. Furthermore, we investigate the robustness of the proposed method against attacks using extended attack scenarios based on the scenarios employed in the Voice Privacy Challenge. Our experimental results confirmed that the proposed method maintains privacy protection performance for speaker concealment, even under more powerful attack scenarios not considered in prior work.
\end{abstract}

\section{Introduction}
Recently, deep learning-based speech-processing systems have commonly been used with smartphones, home assistants, and other edge devices. Such devices have strict computational resource constraints that make it challenging to run inference locally. To address this limitation, the speech-processing systems on these edge devices are typically implemented by offloading the computations of deep learning models to cloud servers. However, the cloud-based systems require sharing speech data with cloud servers. Speech data contains not only the utterance content but also personal identifying information, including language, age, gender, and speaker identity~\cite{metze2007comparison, fan2021exploringwav2vec20speaker}. The personal information embedded in speech data raises significant global privacy concerns. Therefore, under the General Data Protection Regulation (GDPR), an international data protection law, audio data is subject to privacy regulations~\cite{gdpr}. 

Against this background, speech privacy issues have received significant focus~\cite{kai2021lightweight,patino2021speakeranonymisationusingmcadams,Gharib_2024,singh2024voice,tayebi2024addressing}. For example, the Voice Privacy Challenge (VPC), an international competition aimed at advancing speech anonymization technologies, serves as a key initiative within these ongoing efforts~\cite{voiceprivacy2020,voiceprivacy2022,voiceprivacy2024}. The major speaker anonymization approaches involve voice conversion-based techniques, which are employed as the core technology in many methods submitted to VPC~\cite{yao2025npuntuvoiceprivacy2024, xinyuan2024hltcoejhusubmissionvoice, akti24_spsc}. Such speech anonymization methods aim to conceal the speaker identity while preserving the linguistic content in an utterance. 

An encryption-based speech privacy-preserving method with a random orthogonal matrix as a secret key, distinct from the voice conversion-based methods, has been proposed to protect both the speaker identity information and the linguistic content of an utterance~\cite{shoko2024speech}. This method enables deep learning model inference while concealing both speaker identity and speech content. However, this method has limited resistance to inference attacks on speech content and speaker identity. It also cannot be applied to mainstream speech-processing self-supervised learning (SSL) models due to restrictions on compatible model architectures. Furthermore, existing evaluations have only considered attackers without access to encryption systems, indicating that more sophisticated adversaries are needed for comprehensive assessments of speech privacy. 

In this work, we propose a method that enhances the attack resistance of the conventional encryption-based speech privacy-preserving technique by employing multiple random orthogonal matrices as secret keys. We also introduce an approach to relax the model constraints of the conventional method. This approach enables the application of our method to a broader range of deep learning models, including mainstream SSL models. Furthermore, we investigate the robustness of the proposed method against attacks by proposing extended attack scenarios. These scenarios include assessments against sophisticated adversaries with access to the encryption system, inspired by the VPC attack scenario. The experimental results confirmed that the proposed method can be applied to automatic speech recognition (ASR) and automatic speaker verification (ASV) models that use a widely used SSL model as their frontend. Furthermore, we found that the proposed method maintains attack resistance, particularly for speaker identity concealment, even under these more sophisticated attack scenarios.

The contributions of our work are as follows:
\begin{enumerate}
  \item \textbf{Enhanced attack resistance}: We propose a method that employs multiple random orthogonal matrices as secret keys to enhance the attack resistance of the conventional encryption-based speech privacy-preserving method.
  \item \textbf{Broader model applicability}: The proposed method also includes a function that relaxes the model constraints of the conventional method. The function enables the proposed method to adapt to a broader range of deep learning models, including mainstream SSL models, without requiring retraining.
  \item \textbf{Extended evaluation framework}: We advocate comprehensive evaluations using two attack scenarios extended from the scenario employed in VPC, including assessments against sophisticated adversaries with access to the encryption system.
\end{enumerate}

\section{Privacy-Preserving Scenario} \label{sec:pp_scenario}

\begin{figure}[tb]
    \centering
    \includegraphics[width=\linewidth]{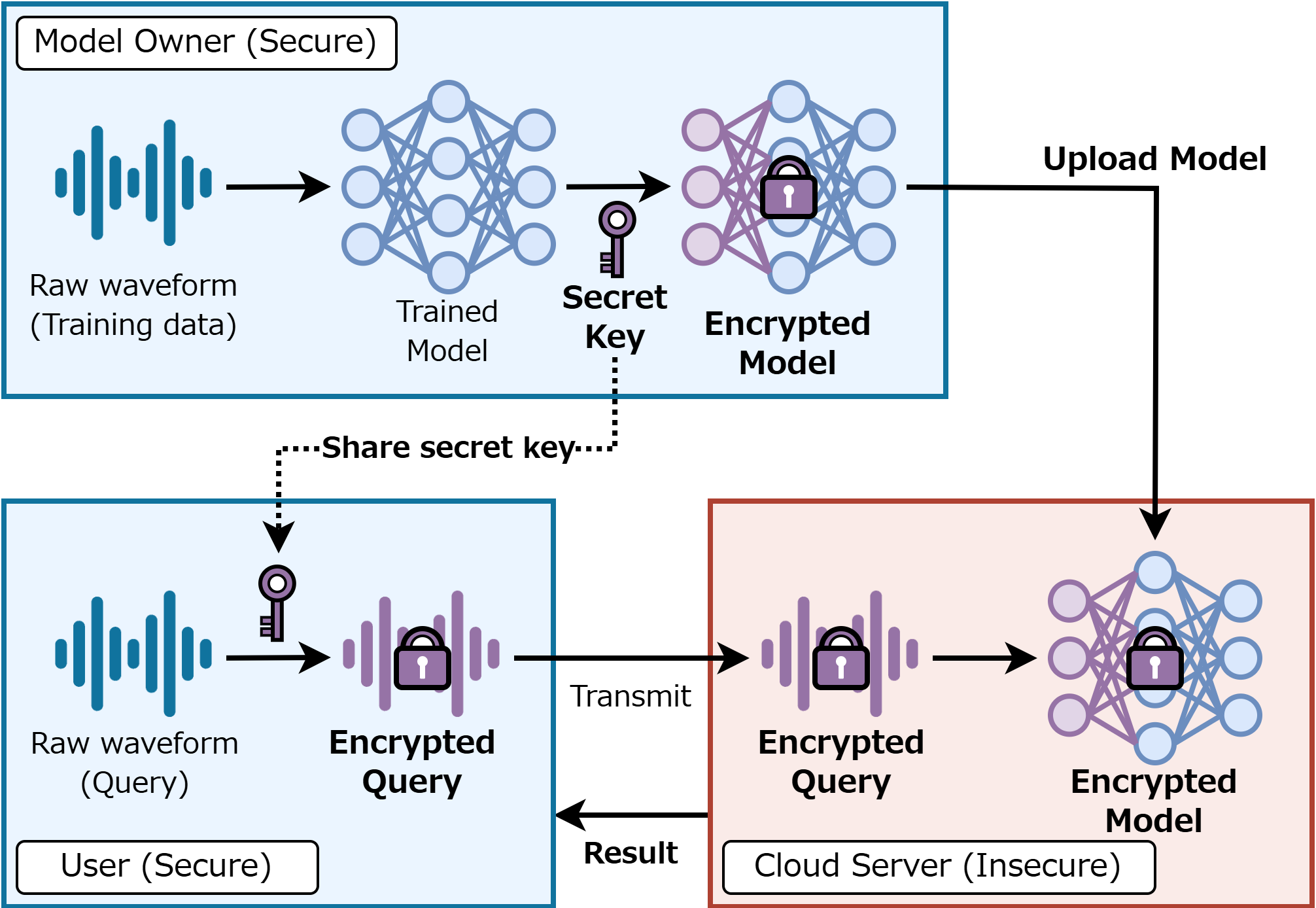}
    \caption{Privacy-Preserving Scenario of the encryption-based methods. The model owner trains a model on unencrypted speech data, encrypts the model using a secret key, and deploys the encrypted model to the cloud server. Users encrypt their speech queries using the shared secret key before transmission. The cloud performs inference on encrypted queries using the encrypted model, ensuring speech privacy without decryption.}
    \vspace{-4mm}
    \label{fig:pp_scenario}
\end{figure}

This section explains the privacy protection scenarios of the encryption-based methods.
Figure~\ref{fig:pp_scenario} shows an overview of the scenario, which consists of three domains: model owner, user, and cloud server. The cloud-based model performs inference on the encrypted speech to obtain results on speech processing tasks while protecting user speech privacy. The workflow operates as follows: The model owner trains the model using unencrypted speech datasets, encrypts the model using a secret key, and deploys it to the cloud. The secret key is then shared with an authorized user. The user encrypts query speech using the shared secret key and transmits it to the cloud. The cloud performs inference using the encrypted model without decrypting the query, thereby preventing third parties from accessing the original speech without the secret key. This approach ensures user privacy by making it difficult to infer information from the encrypted queries.

\section{Proposed method} \label{sec:method}
This section describes the proposed method for enhancing the attack resistance of the conventional encryption-based method and enabling its application to a broader range of deep learning models without retraining.

\subsection{Multiple random orthogonal matrices as secret keys for speech privacy preservation}\label{subsec:multi_mat_key}
To enhance the attack resistance of the conventional method, we propose an encryption-based technique that utilizes multiple random orthogonal matrices as secret keys. This approach encompasses the conventional method in which only one matrix is used as a secret key. The following sections explain the speech encryption and model encryption processes of the proposed method in the scenario described in Section~\ref{sec:pp_scenario}. 

\subsubsection{Speech encryption} \label{subsubsec:speech_enc}
Speech encryption is performed by dividing a one-dimensional speech waveform $\bm{X}$ into blocks $\bm{X}_{i}$ of size $M$ and multiplying each block $\bm{X}_{i}$ by an $M$-dimensional random orthogonal matrix. The speech waveform $\bm{X}$, divided into blocks, is represented as follows:
\begin{equation}\label{eq:whole}
\bm{X}=
\left[
\bm{X}_{1},  \dots, \bm{X}_{i}, \dots, \bm{X}_{t}
\right],
\end{equation}
where $t$ is the total number of blocks.
The secret key $\bm{K}_\text{mult}$ contains $N$ random orthogonal matrices of dimension $M$, as follows:
\begin{equation}\label{eq:key_p}
\bm{K}_\text{mult}=\{\bm{K}_{1}, \dots, \bm{K}_{n}, \dots,\bm{K}_{N}\}.
\end{equation}
When encrypting speech, each block $\bm{X}_{i}$ is encrypted using a random orthogonal matrix $\bm{K}_{n}$ from $\bm{K}_\text{mult}$, and the encryption process is as follows:
\begin{equation}\label{eq:enc2_p}
\bm{X}_{i}^{(\bm{K}_{n})} = \bm{X}_{i} \bm{K}_{n}, \quad n = i\bmod N.
\end{equation}
The random orthogonal matrix $\bm{K}_{n}$ to be applied to the block $\bm{X}_{i}$ is selected based on the block index $i$, such that $n=i\bmod N$.
Encrypted queries sent to the cloud are formed by concatenating the encrypted blocks, as follows:
\begin{equation}\label{eq:enc_p}
\bm{X}^{(\bm{K}_\text{mult})}=
\left[
\bm{X}_{1}^{(\bm{K}_{1})}, \dots, \bm{X}_{N}^{(\bm{K}_{N})}, \bm{X}_{N+1}^{(\bm{K}_{1})}, \dots, \bm{X}_{t}^{(\bm{K}_{t\bmod N})}
\right].
\end{equation}
The encrypted speech makes it difficult even for humans to recognize linguistic content and speaker identity.

\subsubsection{Model encryption} \label{subsubsec:model_enc}
Model encryption is a preprocessing step that modifies the model to enable correct prediction from encrypted queries. Similar to the conventional method, this method targets models that directly process one-dimensional speech waveforms as input. Here, we assume that the first layer of the model is a 1D convolutional layer with equal kernel size and stride. The dimension of the random orthogonal matrices in secret key $\bm{K}_\text{mult}$ must be equal to the kernel size. Consider a simplified convolutional layer with kernel size $M$, stride $S$, number of output channels 1, and no bias term. The kernel $\bm{E}$ of the convolutional layer is expressed as follows:
\begin{equation}\label{eq:conv_kernel}
    \bm{E}=\left[e_{0}, \dots, e_{k}, \dots, e_{M-1}\right]^\top.
\end{equation}
Model encryption is performed by multiplying the kernel of the first layer by the transpose $\bm{K}_{n}^\top$ of the matrix applied to the corresponding speech waveform block $\bm{X}_{i}$. This allows the first layer to cancel out the matrix $\bm{K}_{n}$ applied to block $\bm{X}_{i}$, making the first layer output equivalent to the unencrypted case. The model encryption process is as follows:
\begin{equation}\label{eq:enc_kernel}
\bm{E}^{(\bm{K}_{n})} = \bm{K}_{n}^\top \bm{E}.
\end{equation}
Since $N$ encrypted kernels are prepared, the first layer branches into $N$ encrypted convolutional layers, with each branch processing blocks encrypted with the corresponding $\bm{K}_{n}$. This approach requires routing encrypted speech blocks to different convolutional layers based on the corresponding matrix $\bm{K}_{n}^\top$ applied to each kernel. While this changes the model structure, it eliminates the need for retraining since the encrypted kernels are computed from the pre-trained model and the secret key. When using different secret keys, the model can be re-encrypted by applying new keys to the original unencrypted model without retraining.

\subsection{Mitigating Model Restrictions} \label{sec:proposed_limit}
\begin{figure}[tb]
    \centering
    \includegraphics[width=60mm]{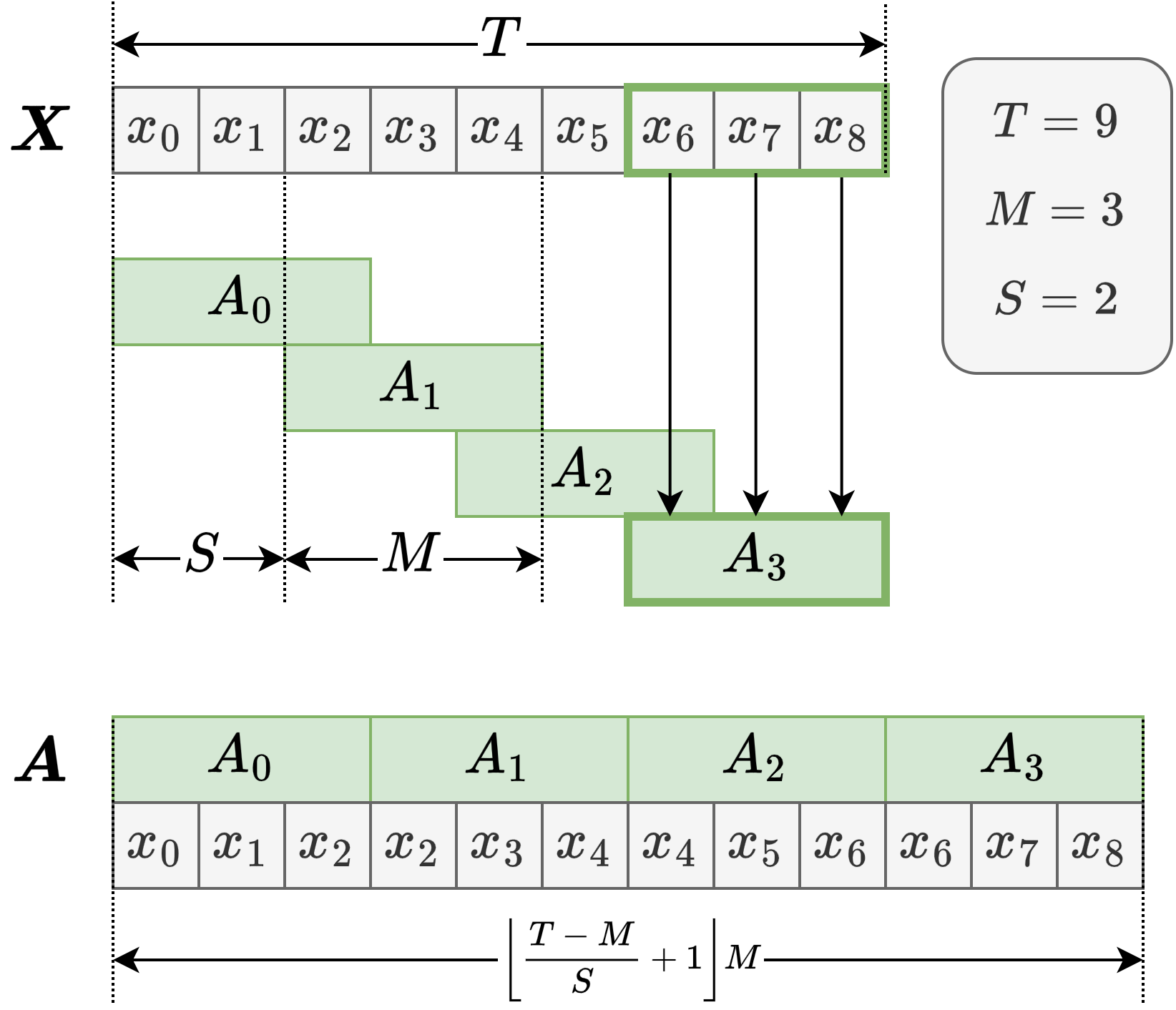}
    \caption{Block partitioning method for handling convolutional layers with stride $S < M$. The input waveform $\bm{X}$ is divided into overlapping blocks $\bm{A}_i$ of size $M$, with consecutive blocks sharing $M-S$ samples. This transformation enables the application of encryption to models where the kernel size and stride differ as shown for $M=3$ and $S=2$.}
    \vspace{-4mm}
    \label{fig:block_proposed}
\end{figure}

As explained in Section~\ref{subsubsec:model_enc}, the conventional encryption-based speech privacy preservation method requires that the kernel size and stride of the model's first 1D convolutional layer be equal to each other. This constraint significantly limited the applicability of conventional methods to models. In this section, we explain how the proposed method, through a modification to the waveform block partitioning method, can relax the constraints on its applicability to deep learning models. This approach enables correct computation during the model encryption, even when the stride of the first convolutional layer does not match the kernel size. Similar to Section~\ref{subsubsec:model_enc}, consider a 1D convolutional layer with kernel size $M$, stride $S$, and no bias term. The kernel of this convolutional layer is expressed as Equation~(\ref{eq:conv_kernel}). The input $\bm{Y}$ and output $\bm{Z}$ of the convolutional layer are expressed as follows:
\begin{equation}\label{eq:input_x}
    \bm{Y}=\left[y_0, y_1, \dots, y_{T-1}\right],
\end{equation}
\begin{equation}\label{eq:output_y}
    \bm{Z}=\left[z_0, \dots, z_i, \dots, z_{L-1}\right],
\end{equation}
where $T$ denotes the length of the input to the convolutional layer, $L$ denotes the length of its output, and $L=\lfloor \frac{T-M}{S}+1 \rfloor$. Under the condition that the kernel size and stride satisfy $S < M$, we assume that the convolutional layer can be computed as an inner product between the block and the kernel. Each element $z_i$ of the convolutional layer output is expressed as follows:
\begin{equation}\label{eq:simple_conv}
    z_i = \sum_{k=0}^{M-1} e_k y_{Si+k} = A_i \bm{E},
\end{equation}
where $A_i = [y_{Si}, y_{Si+1}, \dots, y_{Si+(M-1)}]$, and $A_i$ denotes a vector consisting of elements from the input $\bm{Y}$ that are referenced when computing the output element $z_i$ of the convolutional layer. The array $\bm{A}$ is then formed by extracting and concatenating these vectors $A_i$ ($A_0$ to $A_{L-1}$) from the input audio $\bm{Y}$. Figure~\ref{fig:block_proposed} illustrates this construction of $\bm{A}$ for the case where $M=3$ and $S=2$. As depicted, when the stride S is smaller than the kernel size $M$, consecutive vectors $A_i$ share $M-S$ overlapping samples from the input $\bm{Y}$. This structured concatenation of overlapping blocks $A_i$ is crucial for enabling the transformation of a convolutional layer with $S<M$ into an equivalent one where the effective stride is $S=M$ when $\bm{A}$ is used as input. Here, we consider modifying the convolutional layer to use $\bm{A}$ as input to the model. By changing the stride to $S=M$ without changing the kernel size, when $\bm{A}$ is provided as input, an output that matches the output of the original convolutional layer can be obtained.
The length $L_A$ of the input to the modified convolutional layer becomes as follows:
\begin{equation}\label{eq:output_length}
    L_A = \bigg\lfloor \frac{T-M}{S}+1 \bigg\rfloor M.
\end{equation}
As illustrated in Figure~\ref{fig:block_proposed}, $L_A$ becomes longer than the original input.
By utilizing this method, a convolutional layer with $S<M$ can be modified to a convolutional layer with $S=M$ that performs the same computation. As a result, the proposed method can be applied to models with convolutional layers in which the kernel size and stride differ. Speech encryption can be performed on $\bm{A}$ using the same procedure described in Section~\ref{subsubsec:speech_enc}. For model encryption, we use the same procedure described in Section~\ref{subsubsec:model_enc} but modify the stride of the model's first convolutional layer to $S=M$.

\section{Evaluation of Attack Resistance}
In this section, we propose two attack scenarios, inspired by VPC 2020~\cite{voiceprivacy2020} and VPC 2024~\cite{voiceprivacy2024}, to evaluate the attack resistance of the proposed method. Attacks against encrypted queries aim to infer information, specifically linguistic content and speaker identity,  from original speech embedded within the encrypted queries. Therefore, we extend the VPC scenarios, which originally focused only on speaker identity concealment, to include linguistic content inference attacks.

\subsection{Adversary model} \label{subsec:attack_model}
Adversaries use ASV models and ASR models to infer speaker identity and linguistic content from an utterance, respectively. We assess the attack resistance of the proposed method under scenarios where attackers have access to the following two types of information:
\begin{enumerate}
  \item Encrypted query speech transmitted by users to cloud-based models.
  \item Encryption algorithm used to encrypt query speech.
\end{enumerate}
In contrast, the secret key used for encryption is inaccessible to adversaries. Adversaries leverage the accessible information, along with ASR  and ASV models, to infer linguistic content and speaker identity from encrypted queries. Additionally, adversaries may preprocess encrypted audio or adapt their attack models using the accessible information to improve attack accuracy.

\subsection{Attack scenario 1}
\begin{figure}[tb]
    \centering
    \includegraphics[width=\linewidth]{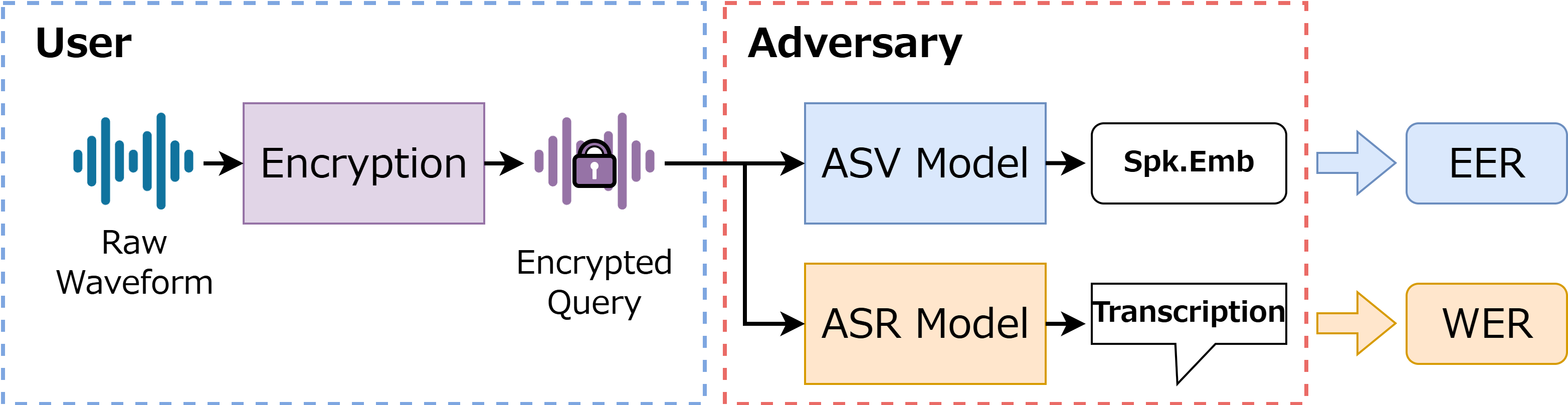}
    \caption{Attack scenario 1: Ignorant attacker with access only to encrypted queries. The adversary intercepts encrypted speech queries and attempts to extract both linguistic content (using ASR models to generate transcriptions for WER evaluation) and speaker identity (using ASV models to create speaker embeddings for EER evaluation) without knowledge of the encryption algorithm or secret key.}
    \vspace{-4mm}
    \label{fig:scenario1}
\end{figure}

Attack scenario 1 assumes that the adversary performs attacks using only information specified in Section~\ref{subsec:attack_model} item (1), i.e., exclusively encrypted queries. This attack scenario is based on the ``ignorant attacker'' scenario from the VPC2020~\cite{vpc2020result}. Figure~\ref{fig:scenario1} shows an overview of attack scenario 1. The adversary obtains encrypted queries transmitted by users and attempts to infer linguistic content and speaker embeddings using pre-trained ASR and ASV models. We use the following two evaluation metrics to assess attack resistance.

\textbf{Linguistic content inference}: Word Error Rate (WER) between the adversary's ASR transcription of the encrypted query and the original transcription indicates how accurately the speech content can be inferred.

\textbf{Speaker identity inference}: Equal Error Rate (EER) is calculated based on the similarity between speaker embeddings extracted by the adversary's ASV model from encrypted queries and speaker embeddings from unencrypted enrollment utterances of the same speaker. 
Higher WER and EER values indicate greater attack resistance for the encryption method.

\subsection{Attack scenario 2}
\begin{figure}[tb]
    \centering
    \includegraphics[width=0.93\linewidth]{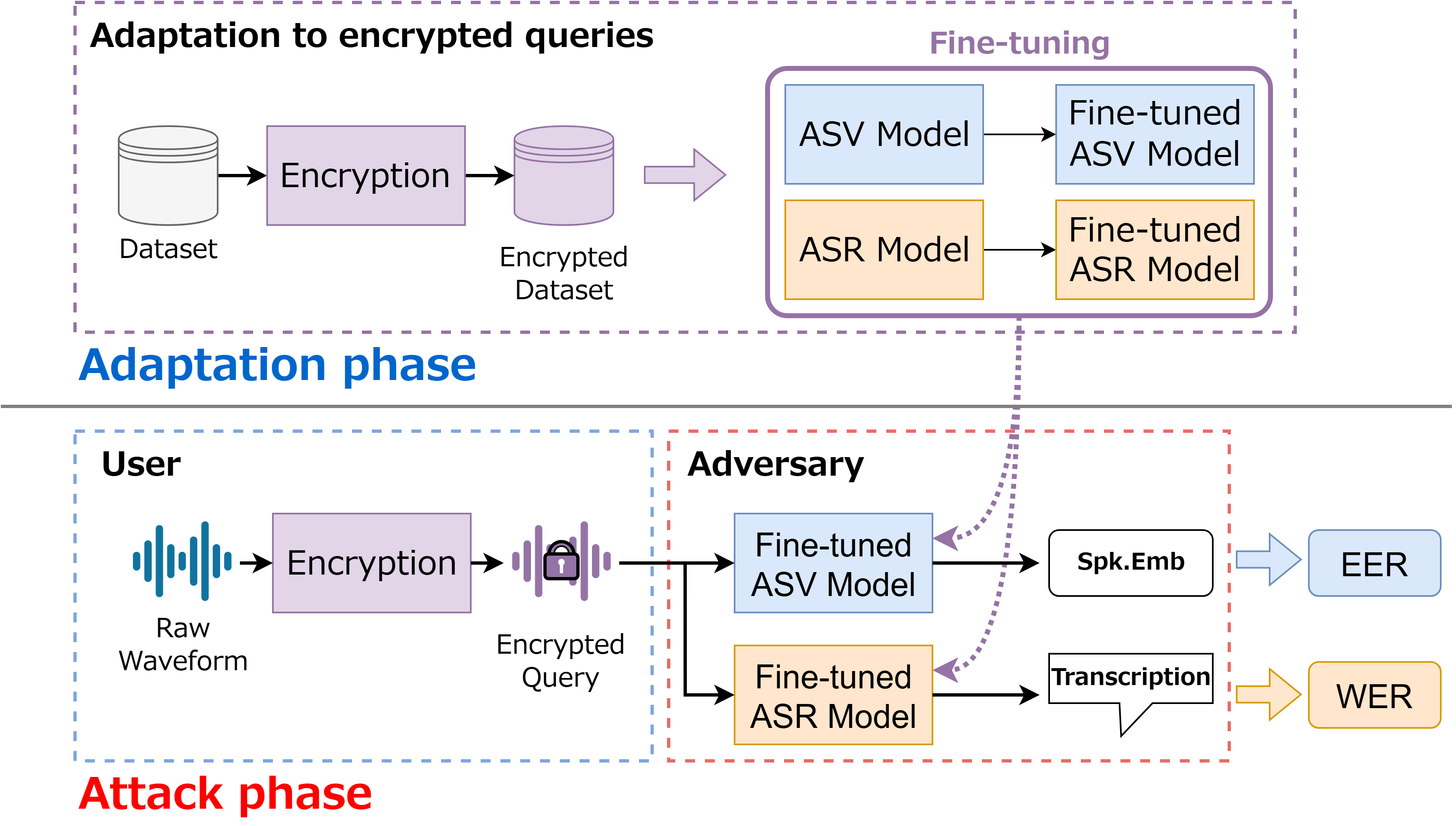}
    \caption{Attack scenario 2: Semi-informed attacker with knowledge of the encryption algorithm. The adversary first adapts to encrypted queries by fine-tuning ASR and ASV models using encrypted datasets generated with the same encryption system (with different key configurations). Then, it uses these fine-tuned models to perform more sophisticated attacks on encrypted queries, inferring linguistic content and speaker identity.}
    \vspace{-4mm}
    \label{fig:scenario2}
\end{figure}

Attack scenario 2 assumes that the adversary performs attacks using both types of information specified in Section~\ref{subsec:attack_model}: (1) encrypted queries and (2) the encryption algorithm. This attack scenario is based on the "semi-informed attacker" scenario from the VPC2024~\cite{voiceprivacy2024}. Figure~\ref{fig:scenario2} shows an overview of attack scenario 2. Unlike in Attack Scenario 1, the adversary in this scenario has knowledge of the encryption algorithm and can generate encrypted datasets using the same encryption system. The adversary fine-tunes ASR and ASV models using these encrypted datasets and uses the adapted models to conduct attacks on encrypted queries. The attack resistance is evaluated using WER and EER metrics. For EER computation, speaker embeddings are extracted from both encrypted query utterances and encrypted enrollment utterances. This approach is used because the ASV model has been adapted to encrypted queries through fine-tuning. Attack Scenario 2 represents a more sophisticated attack than Attack Scenario 1. In this scenario, the adversary leverages models specifically adapted to encrypted queries to improve inference accuracy.

\section{Experiment}
This section presents experiments conducted to validate the proposed encryption method and evaluate its attack resistance. We conducted experiments to validate that the proposed encryption method can be applied to models with different kernel sizes and strides. This was not possible with conventional methods. Additionally, we evaluated the attack resistance of the encryption method through experiments based on the attack scenarios described in Section~\ref{subsec:attack_model}.

\subsection{Pre-trained models}
We utilize wav2vec 2.0 \cite{wav2vec2} Large and wav2vec 2.0 Base as frontends for ASR and ASV models in the experiments. The ASR model backend consists of a 2-layer fully connected layer with 1024 units, while the ASV model uses x-vector \cite{x-vector} for the backend. The first convolutional layer of wav2vec 2.0 has a kernel size of 10 and a stride of 5, representing settings in which previous encryption methods cannot be applied. Both wav2vec 2.0 Large and Base models share the same CNN feature extractor architecture, differing only in the number of Transformer blocks and attention heads. The ASR and ASV models are pre-trained on LibriSpeech \cite{librispeech} and VoxCeleb1 \cite{voxceleb1} datasets, respectively.

For Attack Scenario 2, we fine-tune the pre-trained models using encrypted datasets to create models adapted to encrypted speech for attacks. The encrypted datasets are constructed by encrypting the same datasets used for pre-training, using randomly generated secret keys for each utterance. For ASR model fine-tuning, we use only the train-clean-100 and train-clean-360 subsets of LibriSpeech. Fine-tuning uses learning rates of 1e-4 for the frontend and 5e-3 for the backend.

\subsection{Preprocessing}
As mentioned in Section~\ref{subsec:attack_model}, adversaries can apply preprocessing to encrypted queries within the scope defined by the adversary model. We apply two preprocessing techniques with dual objectives: improving adversary model performance and stabilizing the fine-tuning process using encrypted datasets in Attack Scenario 2.

The first preprocessing is time-scale adjustment of encrypted queries. As described in Section~\ref{sec:proposed_limit}, the length of encrypted queries given by Equation~(\ref{eq:output_length}) is stretched because consecutive blocks $A_i$ and $A_{i+1}$ contain overlapping samples of length $M-S$, as shown in Figure~\ref{fig:block_proposed}. To match the original audio length, we remove overlapping samples between consecutive blocks. The processed result $\tilde{\bm A}$ is expressed using blocks $A_i$ from Equation~(\ref{eq:simple_conv}) as follows: 
\begin{equation}\label{eq:trim_block}
\begin{gathered}
    \tilde{\bm A} = \left[ \tilde A_0, \dots, \tilde A_i, \dots, \tilde A_L \right], \\
    \tilde A_0 = A_0, \\ \tilde A_i = \bigl(A_i[M-S],\,A_i[M-S+1],\,\dots,\,A_i[M-1]\bigr).
\end{gathered}
\end{equation}
The resulting length is expressed as follows: 
\begin{equation}\label{eq:trim_block_length}
\bigl\lvert \tilde{\bm A} \bigr\rvert
= \bigl\lvert \tilde{\bm A}_0 \bigr\rvert
+ \sum_{i=1}^{L-1} \bigl\lvert \tilde{\bm A}_i \bigr\rvert
= M + (L-1)S
= T.
\end{equation}
This achieves length consistency with the original audio.
This preprocessing is used in both Attack Scenarios 1 and 2, and is incorporated into the fine-tuning process for Attack Scenario~2. 

The second preprocessing method is low-pass filtering, applied exclusively in Attack Scenario 1 after the first preprocessing step. A low-pass filter with a cutoff frequency of 4 kHz removes high-frequency components from encrypted query speech before being input to the adversary's model.

\subsection{Experimental results}
\begin{table}[tb]
    \centering
    \caption{Performance of Pre-trained Models Under the Proposed Encryption Method
\newline(No encryption: WER = 1.76[\%], EER = 3.66[\%])} 
    \begin{tabular}{|c||c|c|c|c|} \hline
                    & \multicolumn{2}{|c|}{Correct key} & \multicolumn{2}{|c|}{Incorrect key}\\ \hline
        N           & WER [\%]  & EER [\%]  & WER [\%]  & EER [\%]  \\ \hline\hline
        1           & 1.76      & 3.66    & 67.1      & 30.8      \\
        3           & 1.76      & 3.66    & 94.2      & 34.6      \\
        5           & 1.76      & 3.66    & 96.6      & 36.8      \\ \hline
    \end{tabular}
    \label{tab:result1}
\end{table}

Table \ref{tab:result1} shows the performance of ASR and ASV when encryption is applied to both the pre-trained model and the input speech. $N$ denotes the number of random orthogonal matrices in the secret key. In Table \ref{tab:result1}, when using the correct key, WER and EER are equivalent to the unencrypted case, showing no performance degradation. Here, a correct key means that the speech and model are encrypted with the same secret key. This occurs because the output of the first convolutional layer remains identical to the unencrypted case when both audio and the model share the same secret key, as explained in Section \ref{sec:method}. In contrast, when audio and the model use different secret keys, WER and EER values increase significantly compared to the correct key scenario. These results demonstrate that the proposed method enables encrypted inference without performance degradation for model configurations incompatible with the conventional method. 

Tables \ref{tab:scenario1_result} and \ref{tab:scenario2_result} show the attack resistance evaluation results for Attack Scenarios 1 and 2. Table \ref{tab:scenario1_result} shows that both WER and EER increase with the number of random orthogonal matrices $N$. Specifically, WER increased from 40.6\% ($N=1$) to 97.5\% ($N=9$), and EER increased from 33.1\% to 46.8\%. This indicates that increasing $N$ makes it more difficult to infer linguistic content and speaker identity from encrypted speech, thereby improving privacy protection. Similar trends were observed with low-pass filtering, where WER and EER showed slight decreases, but attack resistance was generally maintained.

Attack Scenario 2 results (Table \ref{tab:scenario2_result}) show the evaluation under stronger attack conditions in which attackers have access to the encryption system. WER increased from 3.66\% ($N=1$) to 11.2\% ($N=9$), and EER increased from 11.4\% to 27.0\%, confirming improved attack resistance with increasing $N$. However, both WER and EER were lower than those in Attack Scenario 1, indicating that attacks using models adapted to encrypted speech are more effective.

These results demonstrate that increasing the number of random orthogonal matrices effectively improves attack resistance in both scenarios. Even against stronger attacks like Attack Scenario 2, a certain level of privacy protection is maintained, particularly for speaker identity concealment.

\begin{table}[tb]
    \centering
    \caption{Evaluation of Attack Resistance in Attack Scenario 1}
    \begin{tabular}{|c||c|c|c|c|} \hline
                    & \multicolumn{2}{|c|}{Without preprocessing} & \multicolumn{2}{|c|}{LPF}\\ \hline
        N           & WER [\%]  & EER [\%]  & WER [\%]  & EER [\%]  \\ \hline\hline
        1           & 40.6      & 33.1    & 38.3      & 30.8      \\
        3           & 90.0      & 44.8    & 87.1      & 45.1      \\
        5           & 93.5      & 46.3    & 90.0      & 46.0      \\
        7           & 96.8      & 46.5    & 94.4      & 46.7      \\
        9           & 97.5      & 46.8    & 94.9      & 47.0      \\ \hline
    \end{tabular}
    \label{tab:scenario1_result}
\end{table}

\begin{table}[tb]
    \centering
    \caption{Evaluation of Attack Resistance in Attack Scenario 2}
    \begin{tabular}{|c||c|c|} \hline
        N           & WER[\%]   & EER[\%]       \\ \hline\hline
        1           & 3.66      & 11.4      \\
        3           & 7.53      & 22.3      \\
        5           & 9.95      & 22.8      \\
        7           & 11.4      & 26.9      \\
        9           & 11.2      & 27.0      \\ \hline
    \end{tabular}
    \label{tab:scenario2_result}
\end{table}

\subsection{Conclusion}
In this work, we proposed an encryption-based method for preserving speech privacy that addresses the limitations of the conventional method. The proposed method enhances attack resistance by utilizing multiple random orthogonal matrices as secret keys and by relaxing model applicability constraints to support a broader range of deep learning models, including mainstream SSL models. To evaluate the attack resistance of the proposed method, we introduced sophisticated attack scenarios that incorporate more challenging attack models and adversaries compared to those typically explored in VPC. The experimental results confirmed that the proposed method enables encryption to be applied to models in which the previous method could not be applied due to architectural constraints. Even under Attack Scenario 2, which assumes more sophisticated  adversaries, the results demonstrated that privacy protection performance for speaker identity concealment is maintained to a reasonable extent. Future work should evaluate the attack resistance of the encryption method using models with various architectures and training data beyond that used in this work.

\section*{Acknowledgment}
This work was supported in part by JSPS KAKENHI (Grant Number JP24K14993), SCAT, and ROIS DS-JOINT (026RP2025) to S. Shiota.

\printbibliography

\end{document}